\documentclass[final,twocolumn,showpacs,preprintnumbers,amsmath,amssymb,superscriptaddress]{revtex4}
\usepackage[utf8]{inputenc}
\usepackage{epsfig}
\usepackage{epstopdf}
\usepackage{amsmath,amssymb,amsthm}
\usepackage{dcolumn}
\usepackage{bm}
\usepackage{graphicx}
\usepackage{subfig}
\usepackage{caption}
\usepackage[]{natbib}
\usepackage{hyperref}

\begin{document}

\title{Monte Carlo tests of Orbital-Free Density Functional Theory} 
\author{D. I. Palade}\email{dragos.palade@inflpr.ro}
\affiliation{National Institute of Laser, Plasma and Radiation Physics,
PO Box MG 36, RO-077125 M\u{a}gurele, Bucharest, Romania}


\begin{abstract}

The relationship between the exact kinetic energy density in a quantum system in the frame of Density Functional Theory and the semiclassical functional expression for the same quantity is investigated. The analysis is performed with Monte Carlo simulations of the Kohn-Sham potentials. We find that the semiclassical form represents the statistical expectation value of the quantum nature. Based on the numerical results, we propose an empirical correction to the existing functional and an associated method to improve the Orbital-Free results.

\end{abstract}

\maketitle

\section{Introduction}
\label{intro}

Density Functional Theory (DFT) \cite{dreizler2011density}\cite{parr1989density} is one of the most successful quantum approaches used nowadays to describe the structure of matter or the dynamics at microscopic level. Currently, the range of applicability spans different scales, from band structures in solid state physics to nuclear matter in nuclei. Between this we find applications in ground-states and dynamics of quantum plasmas and atomic clusters, binding energies and orbitals in molecular physics and electronic structure of atom. The success of the method relies on the fact that it is faster from numerical point of view than other older (but also quantum) methods as Hartree-Fock \cite{baerends1973self} is and, in principle, contains all existing quantum effects. The basic unknown is the density of particles in a system of identical particles and from that a consistent number of observables can be computed. 

Despite being so popular, the theory is far from being complete regarding its applicative power. It is a correct theory, conceptually speaking, but involves an unknown, density dependent mean field potential that contains a large part of pure quantum effects, i.e. exchange-correlations effects \cite{kohn1965self}. Obviously, this term can be neglected only in the classical limit where the number of particles and the length scale becomes very large, therefore, remains crucial for mesoscopic, smaller systems or strongly correlated ones. Much of the world's effort in developing DFT is now concentrated on finding better and better functionals to describe this type of effects, for general systems of for specific ones. 

On a different direction, there is some (small) hope that an universal functional of density for the kinetic energy (the so called Orbital-Free DFT \cite{chen2008orbital}) in a system of Kohn-Sham (KS) \cite{kohn1965self} non-interacting particles will be found. This type of description allows one to write a full density dependent energy functional (for the ground state, or thermodynamic equilibrium) and minimize it under the constrain of constant number of particle. The result would be an Euler-Lagrange equation with the only unknown, the density of particles. This equation could be solved many times faster than the actual Kohn-Sham equations which scale with $O(N^3)$ where $N$ is the number of orbitals involved. Having this kind of functional is not just a matter of being faster in computation but sometimes becomes a necessity. For example, very large molecules, quantum plasma, clusters, quantum dots, etc. which can contain $10^3-10^4$ atoms, are impossible to simulate fully with the KS method, since they can reach a number of electrons of order $10^{5-6}$, or even more, a task far from nowadays computing possibilities.  

Strangely enough, the first genuine DFT (and Orbital Free) appeared almost 40 years before the KS achievement, with the work of L. H. Thomas and E. Fermi \cite{thomas1927calculation}, who developed such a functional in 1927 soon after Schrodinger's equation. But their result is applicable only in cases with very slow varying density of particles. Moreover, it is wrong due to Teller's theorem \cite{jahn1937stability} since molecules do not bind. Later \cite{weizsacker1935theorie}, von Weisecker corrected this functional with an additional term. Currently, there are a series of results regarding the so called gradient expansion of the kinetic energy density in term of probability density, based on the Bloch density matrix. The derivation involves a Wigner-Kirkwood expansion \cite{kirkwood1933quantum} in powers of $\hbar$ and is limited usually to the fourth order. Therefore, this is a semiclassical approximation. An extensive review of the orbital-free approximations can be found in \cite{wang2002orbital} \cite{ligneres2005introduction}, \cite{1973quantum}.

During decades, this functional have been used in different system \cite{brack1985selfconsistent} \cite{iniguez1986density} and different forms, usually with good results at least for the gross properties and trends, since the physical cases were nicely selected, but in realistic ones, as atoms are, the results are not in the acceptable region of error. Of course, the main problem would be to find better approximations or, if it is possible (since there is a question regarding the $N$ representablity \cite{ayers2005generalized}), the true functional. In this paper, our goal is not to solve this mysterious problem, but to investigate how close is the semiclassical form to the quantum reality. All semiclassical expressions for the density of kinetic energy are thought to be an average description in the spatial sense of the quantum reality, due to the fact that the quantum shell effects are not reproduced. Still, can they thought to be an average over the possible quantum systems or, in other words, is the semiclassical result a moment expansion of the quantum DFT ensembles? For this we will focus on the simplified case of a non-relativistic, no spin, $1D$ and ground state system of interacting particles.

The paper is structured as it follows: first, a short description of the DFT and KS method will be given, then the semiclassical functional will be presented. Further, a Monte-Carlo method will be used to describe the statistical relationship between the exact and the approximated version constructing random KS orbitals from random KS potentials.

\section{Theory}
\label{theory}
\subsection{Density Functional Theory}
\label{DFT}

Usually, the interest in the structure of matter is related to electrons or nucleons, so, we will consider a system of identical particles (in particular fermions, even though the spin is not explicitly taken into account) in the non-relativistic limit subject to a two-body interactions of the general form $u(|\vec{r}_1-\vec{r}_2|)$ in the position representation. Also, we denote the state of the system as $|\Psi\rangle$, the associated field operator $\psi(\vec{r})$ and an external field (which in a lot of cases is electrostatic and created by the Coulomb attraction of nuclei) $v_{ext}(\vec{r})$. Therefore, the Hamiltonian operator and the corresponding energy can be written as $H=T+V+U$, where:

\begin{eqnarray}
T=-\frac{\hbar^2}{2m}\int dr^3\psi^\dagger(\vec{r})\nabla^2\psi(\vec{r})\\
V=\int dr^3\psi^\dagger(\vec{r})v(\vec{r})\psi(\vec{r})\\
U=\frac{1}{2}\int dr^3dr'^3\psi^\dagger(\vec{r})\psi^\dagger(\vec{r'})u(\vec{r},\vec{r}')\psi(\vec{r'})\psi(\vec{r})
\end{eqnarray}

In the seminal paper of Hohenberg and Kohn \cite{hohenberg1964inhomogeneous} they proved two theorems that brought DFT on a solid mathematical ground. First one shows that there is an unique mapping between the external potential $v(\vec{r})$ and the density of probability (the diagonalized density matrix in coordinate representation) $\rho(\vec{r})=\langle\Psi|\psi(\vec{r})^\dagger\psi(\vec{r})|\Psi\rangle$. The second theorem proves that there is an unique functional of density , $E[\rho]$, called energy which is minimized only by the true ground state density and which can be written in our case as: 

$$E[\rho]=\int dr^3v(\vec{r})\rho(\vec{r})+\frac{1}{2}\int dr^3dr'^3\rho(\vec{r})u(\vec{r},\vec{r}')\rho(\vec{r}')+G[\rho]$$

Where $G[\rho]=\int g[\rho] dr^3$ is an universal functional of density containing the kinetic energy and many-particle quantum effects (exchange-correlations). If a formal minimization of $E[\rho]$ is performed under the constrain of constant number of particles $N=\int dr^3\rho(\vec{r})=const$, an Euler-Lagrange equation can be obtained:

\begin{equation}
\label{Euler}
\frac{\delta G[\rho]}{\delta \rho}+v(\vec{r})+\int dr'^3\rho(\vec{r}')u(\vec{r},\vec{r}')=\mu
\end{equation} 

Where $\mu=\partial E/\partial N$ is interpreted as chemical potential. Of course, the explicit form of $G[\rho]$ being unknown, this equation seems to be of no practical use. But, one year after HK theorems, Kohn and Sham \cite{kohn1965self} have developed a method based on a set of fictive non-interacting particles described by the set of wave-functions $\phi_i(\vec{r})$ splitting the $G$ functional in a sum of kinetic energies and a term of so called exchange-correlation energy \cite{van1994exchange} \cite{mcguire2005electron}. With this, the DFT can be used as the KS equations :

\begin{eqnarray}
\label{KS}
H_{KS}\phi_j(\vec{r})=[-\frac{\hbar^2}{2m}\nabla^2+V_{KS}(\vec{r})]\phi_j(\vec{r})=\varepsilon_j\phi_j(\vec{r})\\
V_{KS}(\vec{r})=v_H(\vec{r})+v_{ext}(\vec{r})+v_{xc}(\vec{r})
\end{eqnarray}

The terms in the KS potential can be easily identified as external potential, hartree potential (in the case of electrons where the interaction is Columbian) and exchange-correlation potential $v_{xc}=\delta E_{xc}/\delta \rho$. The Hartree potential can be written in terms of density $\rho(\vec{r})=\sum_{j=1}^N|\phi_j(\vec{r})|^2$ as $v_H(\vec{r})=\int dr'^3\rho(\vec{r}')u(\vec{r},\vec{r}')$. As said before, a lot of work has been done to construct good approximations for $v_{xc}$ starting from LDA \cite{parr1994density} up to complex functionals \cite{perdew1992accurate}. This is the form in which almost all DFT calculations are carried out in the present. An important aspect is that the single guaranteed result is the density of particles and the highest energy of the occupied levels \cite{PhysRevB.56.16021}. All other interpretations in terms of single particles solutions $\phi_j$ are unrealistic (although many times used) since the theory states that this are not true single particle wave functions, just a set of auxiliary mathematical entities with the soul purpose of mathematical tool.

Obviously, solving eq. (\ref{KS}) requires to solve iteratively $N$ self-consistent Schrodinger equations which in $3D$ can be a demanding computational task, a lot of time being required to construct the effective KS potential from the density. Typically, such simulations are performed on clusters of processors and the codes are heavily parallelized for large systems. Still, as the number of particle increases and there is no reduction of dimensionality by symmetry, one can expect for days of computation time. And after all of that, there is the problem of dynamics which involves the propagation of a set of orbitals through time \cite{gross1990time} which, of course makes any simulation even more difficult.

Looking back at the Euler-Lagrange equation (\ref{Euler}), one could easily see how, if we would know the $g$ functional we would simplify the numerical treatment by orders of magnitude since we don't need to diagonalize a hamiltonian which is self consistent with the solution, just to solve a non-local equation in one single unknown, e.g. the density. This is the essence of Orbital-Free DFT: to find the $g$ functional.

\subsection{Gradient expansion of Orbital Free Functional}
\label{gradient}

Similarly with the KS method, one can start from eq. (\ref{Euler}) and split the functional $g[\rho]=\tau[\rho]+v_{xc}[\rho]$, where $\tau$ is the density of kinetic energy for the pseudo-particles $\{\phi_i\}$. Therefore, the later can be linked to $\tau$ by:

\begin{equation}
\label{true}
\tau(\vec{r})=-\frac{\hbar^2}{2m}\sum_{j=1}^N\phi_j^*(\vec{r})\nabla^2\phi_j(\vec{r})
\end{equation}

But keeping in mind that in the calculation of the total kinetic energy for finite systems the domain of integration is equal with the entire ($\mathbb{R}^d$) space, using Green's theorem and the condition of null wave function at infinity, one can find two other equally valid functionals:

\[\tau_1(\vec{r})=\frac{\hbar^2}{2m}\sum_{j=1}^N|\nabla\phi_j(\vec{r})|^2\]
\[\zeta(\vec{r})=\frac{\tau(\vec{r})+\tau_1(\vec{r})}{2}\]

Linked by $\tau_1(\vec{r})=\tau(\vec{r})+\hbar^2/8m\nabla^2\rho(\vec{r})$. Same logic can be used for infinite systems were periodic boundary conditions must be employed. The goal of OFDFT becomes to obtain the functional relation between $\tau$ (or $\tau_1$, $\zeta$) and density $\rho$.
In the paper of L.H. Thomas \cite{thomas1927calculation} an approximation for the density of kinetic energy was derived starting from the non-interacting free electron gas: $T_{TF}=\kappa_0\int\rho^{5/3}dr^3$. This expression was the first DFT calculations ever done, but soon it was proven that the results are far from experimental data and moreover, due to Teller \cite{jahn1937stability} it is physically incorrect since the molecule do not bind. Also the asymptotic form of the resulting density in a Thomas Fermi atoms was incorrect. A first correction was introduced by von Weiszacker \cite{weizsacker1935theorie} who added a supplementary term $T_{W}=\frac{\hbar^2}{2m}\int\rho^{1/2}\nabla^2\rho^{1/2}dr^3$ valid only for one particle but which could correct the binding problem.

A solid expression and proof for such a functional can be obtained in the frame of Bloch density matrix with a Kirkwood-Wigner expansion \cite{kirkwood1933quantum}, \cite{wigner1932quantum}. The main idea is to expand $C(\vec{r},\vec{r}',\beta)=\sum_{i=1}^N\phi_i^*(\vec{r}')\phi_i(\vec{r})e^{-\beta\varepsilon_i}$, the Bloch density matrix, around its value obtained in the frame of TF approximation:

\[C_0(\vec{r},\vec{r}',\beta)=(\frac{m}{2\pi\hbar^2\beta})^{3/2}\exp[-\beta V \frac{\vec{r}+\vec{r}'}{2}-\frac{m}{2\hbar^2\beta}(\vec{r}-\vec{r}')^2]\]

\begin{equation}
\label{expansion}
C(\vec{r},\vec{r}',\beta)=C_0(\vec{r},\vec{r}',\beta)(1+\hbar\chi_1+\hbar^2\chi_2+...)
\end{equation}

Where $\beta=(k_BT)^{-1}$. We skip the entire set of calculation since it can be found in \cite{1973quantum}, \cite{jennings1976extended} and remind just the basic steps: the density matrix $\rho(\vec{r},\vec{r}')$ is the inverse Laplace Transform of $C(\vec{r},\vec{r}',\beta)$ and the density of kinetic energy $\tau(\vec{r})=\nabla_{\vec{r}}\nabla_{\vec{r}'}\rho(\vec{r},\vec{r}')$. The odd moments in the expansion (\ref{expansion}) have null integrals, therefore, just the even moments contribute to the functional and the relation between moments and density is obtained replacing the resulting potential in the expansion from the TF equation. Finally, the total kinetic energy can be approximated up to the fourth order in $\hbar$ as:

\begin{eqnarray}\nonumber
\label{semiclass}
G[\rho]&=&\frac{\hbar^2}{2m}\int dr^3\{\kappa_0\rho^{5/3}+\kappa_2\frac{(\nabla\rho)^2}{\rho}+\\&& +\kappa_4[8(\frac{\nabla\rho}{\rho})^4-27(\frac{\nabla\rho}{\rho})^2\frac{\nabla^2\rho}{\rho}+24(\frac{\nabla^2\rho}{\rho})^2]
\end{eqnarray}

With $\kappa_0=3/5(3\pi^2)^{2/3}$, $\kappa_2=1/36$, $\kappa_4=(6480(3\pi^2)^{2/3})^{-1}$ for $3D$ systems. 

Terms over the fourth order diverge, thus this is the usual form in which the functional it is used. Moreover, even the terms in $\hbar^2$ can develop divergences at the turning points \cite{bhaduri1977turning} but these can be removed by a selfconsistent calculation \cite{bartel1984semiclassical}. Beside this well constructed result, there are other attempts to obtain such functionals, based on the correct asymptotic behavior of the density and the correct linear response \cite{foley1996further}, but we will not discuss them here. 

For consistency, let us denote further with $\tau$ the true value of the density of kinetic energy and with $\tau_S$ the semiclassical one (the integrand of (\ref{semiclass}) functional).

\subsection{Monte Carlo analysis of random functionals}
\label{Montecarlo}

We want to investigate how good is this semi-classical approximation $\tau_S$ (\ref{semiclass}), up to order four in general application. Previous tests have been performed on specific simplified systems, with harmonic \cite{brack2001simple}, linear, billiard \cite{brack2009closed}, etc., potentials, or in specific systems, as metal clusters \cite{palade2014general}, \cite{palade2014optical} or nuclei \cite{bethe1968thomas}. But, how close can we expect to be the results of $\tau_S$ in an generic unknown external potential? How does the results depend on the magnitude, on the number of particles, etc.?

Since, especially for the ground state, the interest is in systems with confined particles we will keep that in mind when we will construct the potentials. Other than that, our potentials are not restricted in any way.

To answer to the up-fronted question, we should take all the possible potentials, solve the KS eqs (\ref{KS}) and the Euler-Lagrange eq (\ref{Euler}) and compare the resulting densities, or the resulting densities of kinetic energy. But obviously, this is an impossible task, both in principle and in practice given the complexity of a DFT simulation. Therefore, we seek for a method to avoid solving any KS equation, or at least, to avoid solving it self-consistently.

In order to achieve this goal we use \emph{random potentials}. Taking into account the first Hohenberg-Kohn theorem, that there is an unique mapping between the external potential and the density and also, keeping in mind the expression for the effective KS potential in KS equations, we can extend the mapping between the effective potential and the density, in the sense that, for every \emph{known} $v_{KS}$, solving KS equations, we obtain a density $\rho_{KS}$ with which we can construct $v_H$ and $v_{xc}$ and compute the external potential as $v_{ext}=v_{KS}-v_H-v_{xc}$. It would be reasonable to consider that, if in principle, the $v_{KS}$ spans all the possible forms, then, also $v_{ext}$ would span all the possible forms. We do not have any proof for this matter since we don't know if constructing $v_H$ and $v_{xc}$ we would not restrict the domain of $v_{ext}$. Therefore, we will just accept the statement as a reasonable expectation.

Still, a set of problems and questions arise. First of all, how to take into account so much possibilities, given the fact that we can have others particles than electron and therefore the mass can have different values? On the other hand, the scale of the system ($L_0$) can be quite different: in nuclei we have $L_0\sim 1fm$, in atoms and molecules $L_0\sim 1\mathring{A}$ while in non-periodic mesoscopic systems (clusters, quantum plasmas) we can reach $L_0\sim 10 nm$. The magnitude of the potential can be also problematic, since for nuclei is many orders higher than for electrons. This type of dimensional problems will be embedded in a single parameter, scaling the KS equation in spatial characteristic length $\vec{r}\to L_0\vec{r}$ and in magnitude of the potential $V=\alpha\tilde{V}$:

\begin{equation}
[-\tilde{\nabla}^2+\lambda\tilde{V}(\vec{r})]\tilde{\phi}_j(\vec{r})=\tilde{\varepsilon}_j\tilde{\phi}_j(\vec{r})
\end{equation}

With $\lambda=2mL_0^2\alpha/\hbar^2$, and $\tilde{\varepsilon}_j=2mL_0^2\varepsilon_j/\hbar^2$. Both for nucleons or electrons we have an order of magnitude roughly $\lambda\sim 1$, $\tilde{\varepsilon}\sim 1$. Since we are not interested in the energy of the orbitals, we will just keep a coupling constant $\lambda\in [0,100]$ in front of the potential. 

The other question, which is more difficult to deal with, is: how to assure a reasonable form of the effective potential? Well, as we said before, we will focus only on bound states therefore, the potential will be kept always negative. Beside that, given the fact that all its strength is contained in $\lambda$ we will construct only potentials normalized to the unity in their maximum. Also, all realistic systems have smooth potentials (especially due to screening) and even in atoms, it is a standard method to eliminate the singularity of the electrostatic potential from the nucleus with non-singular pseudo-potentials. So we construct an ensemble $S(\lambda,{c_i},P)$ of potentials with the above properties, using a basis of gaussians (ensuring smoothness) coupled with random coefficients $c_i\in [0,1]$:

\[V_{\lambda,C}(\vec{r})=\sum\limits_{i=1}^{\infty}c_i\exp(-\alpha|\vec{r}-\vec{r}_i|^2)\]

The coefficients are constructed from a function of probability $P(c)$. For simplicity, we work in $1D$ with the spatial variable $x\in [-1,1]$. This involves that $x_i\in [-1,1]$ for the gaussian basis. The laplacian is discretized in a finite difference scheme and the eigenvalue problem is solved. The obtained eigenvalues are discarded since we have no interest in them and the result is an ensemble of $\Phi(\lambda,{c_i},N,P)$ orbitals. The number of orbitals taken into account can be chosen to be the lowest existing ones in the limit of $\varepsilon_j<0$ (bound states). For every configuration from $\Phi(\lambda,{c_i},N,P)$ we compute $\rho_{\lambda,C,N,P}(x)$ and $\tau_{\lambda,C,N,P}(x)$.

In principle, if $\tau_S[\rho]\equiv\tau[\rho]$ would be true, the solution of Euler-Lagrange equation (\ref{Euler}) and $\rho_{KS}$ should match exactly. Since this is not the case we write $\tau_S[\rho_{KS}]\equiv\tau[\rho_{KS}]+\eta[\rho_{KS}]$ and investigate the properties of $\eta$ functional, fitting it in powers of $\tau_S$: 

\begin{equation}
\label{interpol}\eta=\sum_{i=1}^{\infty}a_i\tau_S^i
\end{equation}

\section{Results}
\label{result}

\subsection{Statistical results}
\label{statistic}

From numerical point of view we have used a basis of gaussian function with $M=100$ elements and $\alpha=-100$. The discretization was done in $nx=10000$ equal intervals. While for small $\lambda$ this level of accuracy was not necessary, for large values of $\lambda$ and $N$, the high energy orbitals present high oscillating parts, subject to a necessary more refined grid.
\begin{figure}
\resizebox{0.45\textwidth}{!}{\includegraphics{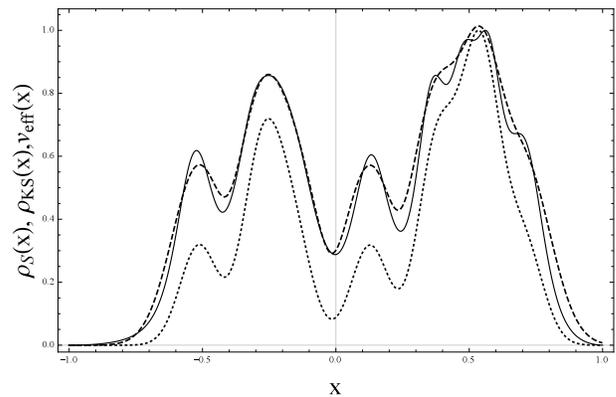}}
\caption{Semi classical density $\rho_S$ (dashed line), DFT density $\rho_{KS}$ (full line) obtained with the same effective potential(dotted line). The number of particles is $9$. All quantities are normalized and the potential $v_{eff}$ is represented as $-v_{eff}$}
\label{fig:1}       
\end{figure}
\begin{figure}
\resizebox{0.45\textwidth}{!}{%
  \includegraphics{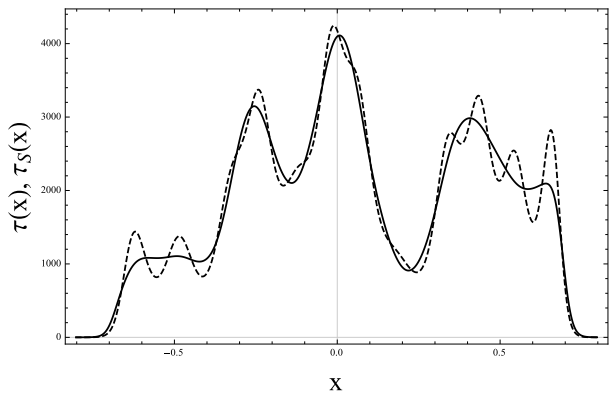}}
\caption{True kinetic energy density $\tau[\rho_{KS}]$ (full line) and the semiclassical valued one $\tau_S[\rho_{KS}]$ (dashed line) under the same effective potential; the later contains false oscillations due to shell effects in the density and $\propto\rho^{3}$ factor in $\tau_S$}
\label{fig:2}       
\end{figure}
As said in the previous section, the coefficient are generated randomly from a distribution function $P(c)$ with the property of being normalized and zero for $c\in \mathbf{R}\backslash[0,1]$. Beside $\lambda$, which controls the strength (or the debt) of the potential, it is very important to generate a variate range of shapes of the potential. Obviously, a random uniform distribution coupled to a large $M$ tends to generate almost constant potential. On the other hand, a normal (Poisson) distribution, tends to generate a localized potential and has also a tail beyond the allowed spatial region. And the list of defects can go on. Therefore, we have performed simulations with $5$ different distributions: normal, arcsine, logitnormal, uniform, U-Qdratic with different parametrization for each and even linear  combinations.

The value of $\lambda$ has been varied over $(0,10^2)$ divided in $10^3$ points. Thus, for every value of $\lambda$ and each distribution function, we have simulated $10^4$ sets of ${c_i}$. Therefore, we have diagonalized $10^7$ hamiltonians, for each one, a variable number of sets $\rho,\tau(\rho), g(\{\phi_j\},\eta(\rho))$, depending on the number of bounded states obtained. For each of them, the (\ref{interpol}) interpolation (with the conjugate gradient method) has been performed to obtain the coefficients ${a_k}$. 

\begin{figure}
\subfloat{\includegraphics[width = 4.5cm]{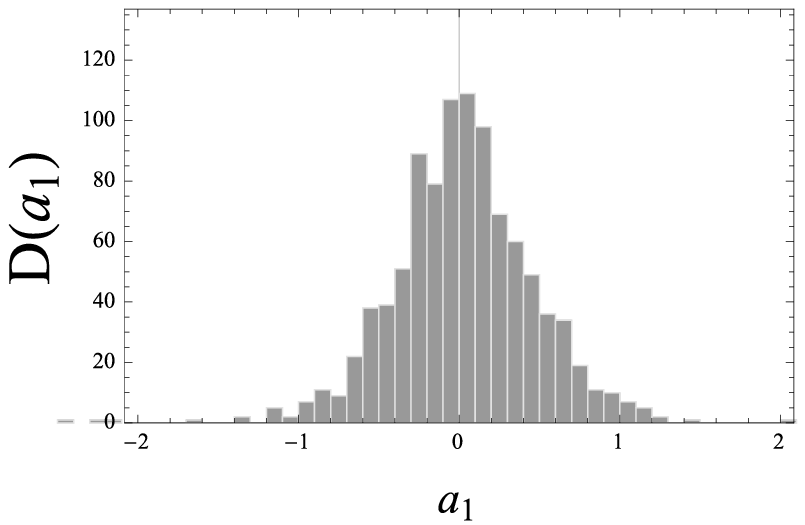}}
\subfloat{\includegraphics[width = 4.5cm]{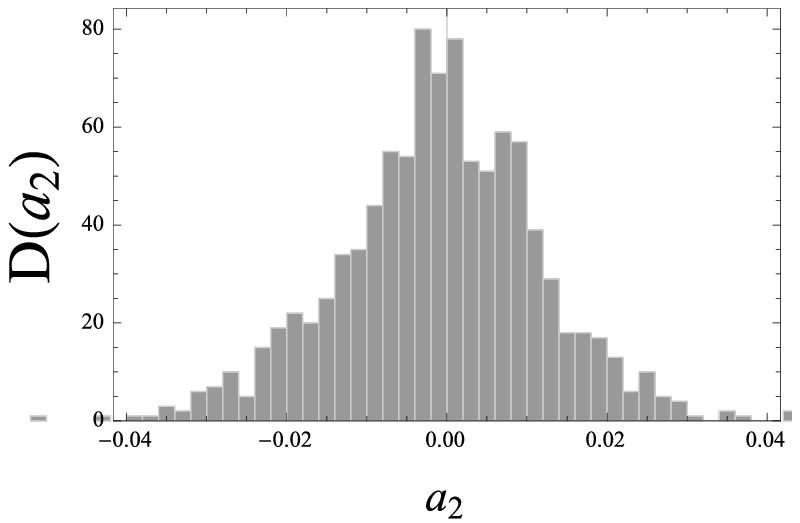}}\\
\subfloat{\includegraphics[width = 4.5cm]{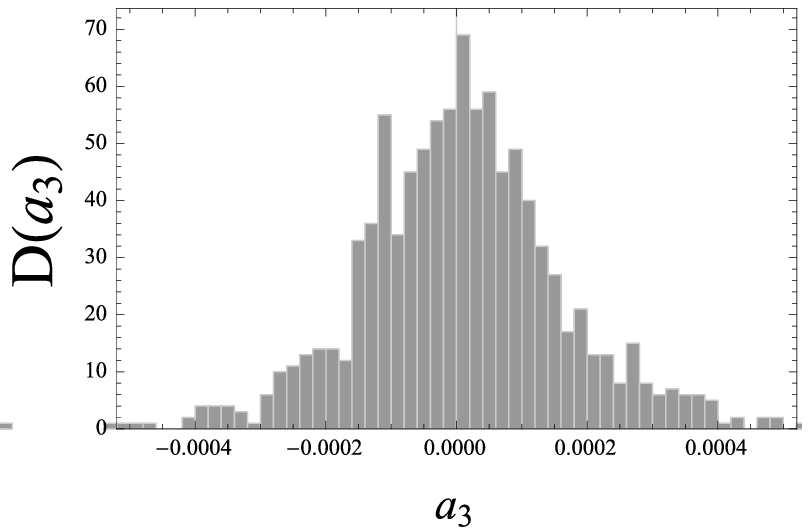}}
\subfloat{\includegraphics[width = 4.5cm]{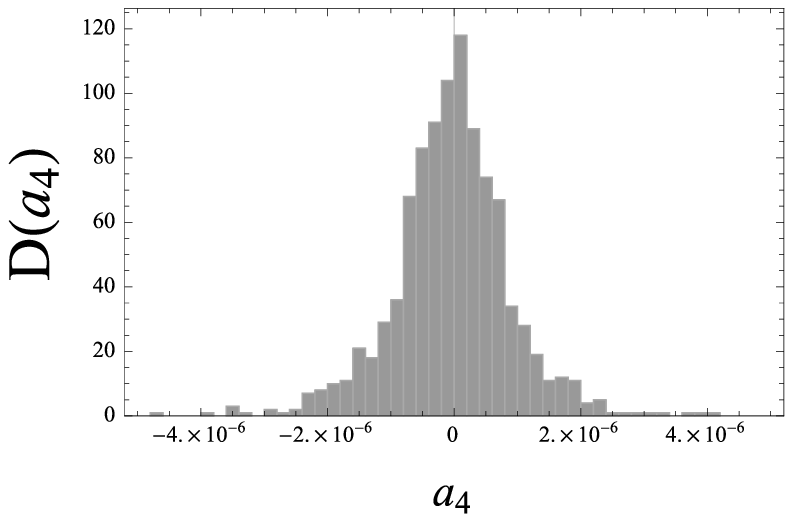}}
\caption{Statistical distribution of coefficients $a_1$,$a_2$,$a_3$,$a_4$ for $N=10$ over all the simulated potentials; the null mean can be observed for each one}
\label{histograms}
\end{figure}

As one can see in Fig. \ref{histograms}, a generic figure for the histogram of the coefficients obtained after $1000$ simulations is plotted. This type of distribution can be seen with any of the random potential used and for any numbers of states involved. That is an approximate gaussian around $0$. This tells us that the semi classical functional $\tau_S$ and the corresponding equation of state is \emph{statistically correct in the quantum world.} 

Still, one could argue that, for electrons let's say, $\lambda$ has very large values only for large scale, close to the bulk domain, where the quantum effects fade away and the semi-classical functional works by definition. Also, in here, large numbers of particles are taken into account and so, they tend to take the lead in the statistic result. Of course, this is true and for that reasons we have also separated the results on the number of particles. Still, the generic gaussian shape is recovered with a larger width.

\subsection{Empirical extension}
\label{empiric}

\begin{figure}
\resizebox{0.5\textwidth}{!}{%
  \includegraphics{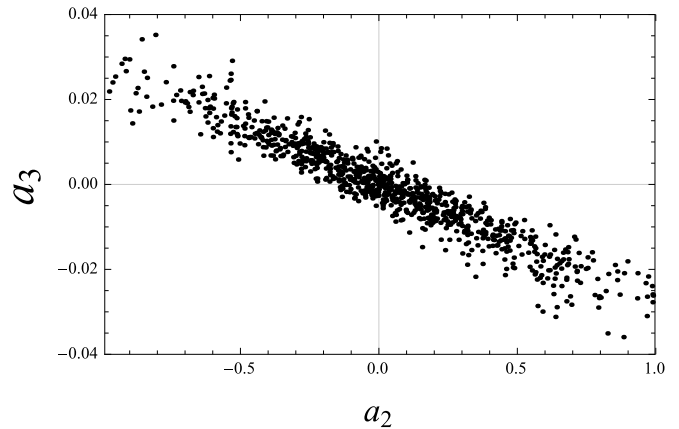}}
\caption{As an example, the statistical relationship over $3000$ of simulations between $a_3$ and $a_2$}
\label{coef}
\end{figure}

Beside the statistical aspect of coefficients, that of being symmetrically distributed around 0, we can search for some dependency between them. More as a consequence of the numerical method used to interpolate the numerical data with the polynomial \ref{interpol}, we find an overall linear dependence between consecutive coefficients, which, obviously, can be approximated by a line. This can be seen in Fig. \ref{coef}.

Moreover, studying numerically the slopes obtained from a least-square fitting with a line of this type of data, we find that can be approximated roughly by recurrent relation as $a_{i-1}=(-2)a_i$. Even if this results is just empirical and possibly connected with the numeric of interpolation, it is to be trusted since the fitting is performed with high accuracy. This results are not $N$ or potential dependent. 
In order to use and test this result, we write the Euler-Lagrange eq \ref{Euler} as an eigenvalue problem adding and subtracting a Bohm potential. The subtraction gives us a Schrodinger like first term, while the addition and the semiclassical expression for $\tau$ are embedded in an effective $w(\vec{r})$ potential:

\begin{eqnarray}
\label{embedded}
[-\frac{\hbar^2}{2m}\nabla^2+w(\vec{r})]\sqrt{\rho(\vec{r})}=\mu\sqrt{\rho(\vec{r})}\\
w(\vec{r})=v_{KS}(\vec{r})+\frac{\hbar}{2m}\frac{\nabla\rho^{1/2}}{\rho^{1/2}}+\frac{\partial\tau_S^{emp}}{\partial\rho}\\
\tau_S^{emp}=\tau_S+a_1 \sum\limits_{i=1}^{\infty}\tau_S^ic_i\\
c_i\approx (-2)^{-i}
\end{eqnarray}

With $\tau_S^{emp}$ is denoted the semi-classical functional with our empirical correction and $\mu$ is the smallest eigenvalue (to obtain the ground state). The choice for an eigenvalue equation is motivated by the possible divergent points in the Bohm potential and the fact that solving an Euler-Lagrange equation in the form \ref{Euler} is more involved numerically due to its high complexity and need of finding an appropriate chemical potential $\mu$ able to maintain the normalization. Otherwise, we must solve eq. \ref{embedded} also in a self-consistent manner as typical Kohn-Sham equations. What is essentially different is the fact that in our effective potential $w(\vec{r})$ we have a term parametric dependent on $a_1$ which is unknown. To avoid this issue, one should choose different values for $a_1$ from a symmetric interval around $0$, but in general $a_1\ll 1$ and solve for each one, the eq \ref{embedded}. From all those solutions, the one which gives the smallest total energy will be chosen. 

Even though we have to solve self-consistently a Schrodinger-like equation several times (let us say $k$ times) to establish which is the best result, still, the method is roughly $N/k$ times faster than KS method which must be solved for $N$ particles.

\begin{figure}
\resizebox{0.5\textwidth}{!}{%
  \includegraphics{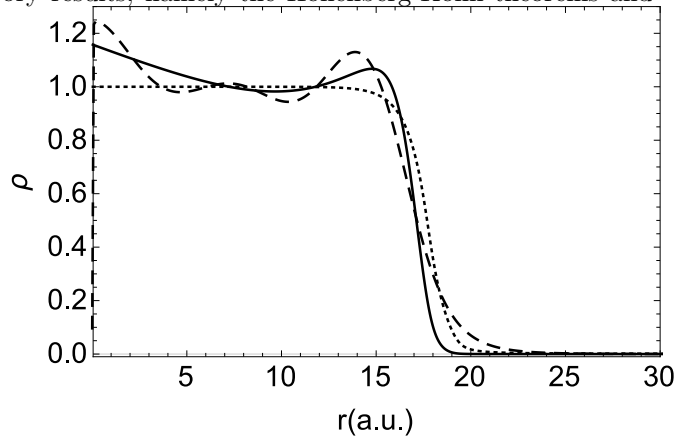}}
\caption{Electron densities in $Na_{92}$ cluster obtained with DFT-LDA (dashed, the shell effects can be observed), Thomas-Fermi (dotted) and the present empirical parametrization (solid line); $a_0\approx 0.02$; The densities are normalized to the TF maximal value}
\label{compar}
\end{figure}

As a test for this method, we apply it in the case of a spherical $Na$ clusters , specifically $Na_{92}$, with the ionic background modeled by the jellium model \cite{brack1993physics}. The results are depicted in Fig. \ref{compar} where the obtained density with Thomas-Fermi method, our parametrization and full DFT-LDA approximations are represented. While no shell effects can be reproduces, and the differences may not be observable in the profiles, the empirical parametrization offers a solution $4$ times more closer (in the squared error $||\rho_S-\rho_{KS}||/||\rho^{emp}_S-\rho_{KS}||\approx 4$ ) to the DFT result. Also the chemical potential is  improved with $30\%$. This is just a basic test for the present proposal capable to validate its mild capabilities. Further tests should be performed on more complicated systems. 

\section*{Conclusions} 

Starting from the classical Density Functional Theory results, namely the Hohenberg-Kohn theorems and the Kohn-Sham method we described the relationship between the pseudo-orbitals and the true density of kinetic energy $\tau$. The later is approximated by a Wigner-Kirkwood expansion around the Thomas-Fermi value of the Bloch density matrix and limited to its semiclassical value $\tau_S$ up to the fourth order. Further, following a set of assumptions on the effective potential we construct an ensemble of random potentials and for each of them we have fitted $\tau$ by a power expansion in $\tau_S$. 

After large number of simulations we gather the coefficients from the above mentioned expansion and found that statistically they fall almost gaussian distributed around the zero value. This allows us to conclude that \emph{the semiclassical functional for the kinetic energy of a system of interacting particles is a statistical mean in the universe of DFT description for the, not only in the spatial sense.} Further, investigating the relationship between coefficients we propose a parametrization for an extended semiclassical functional of the form:

This new form allows one to solve an Orbital-Free problem several times with different values of $a_1$ and retain only the solution with gives the minimal value in energy. While this approach is new and lacks confirmation, local shell effects or asymptotic behaviors in realistic systems, our expectations are that will provide a reasonable step towards the Kohn-Sham results.

\bibliographystyle{unsrt}
\bibliography{bibMonteCarlo}

\end{document}